\documentclass{notices}
\usepackage{amsfonts,amssymb,amsmath,amscd}
\usepackage{graphicx}
\usepackage{caption}
\usepackage{subcaption}

\title{What is Entanglement?}

\author{
  Chon-Fai Kam
  \affil{
    Chon-Fai Kam is a postdoc of Department of Physics at the State University at Buffalo. His email address is dubussygauss@gmail.com.
    }
  \and
  Zhong-Tang Wu
  \affil{
    Zhong-Tang Wu is a PhD student of Department of Applied Physics at Tunghai University of Taiwan. His email address is mpcsphilo@gmail.com.
   }
}

\begin{document}

\maketitle

\section{}
Entanglement, a puzzle since Einstein's time, has become increasingly crucial with the rise of quantum computation. But what exactly is it? Entanglement can be precisely defined, but only negatively. As an axiom, the Hilbert space of a composite system of two sub-systems with Hilbert spaces $H_A$ and $H_b$, is given by $H_A\otimes H_B$. If a state of the composite system can be written as
\begin{equation}
    |\psi\rangle_{AB}=|\psi\rangle_A\otimes |\psi\rangle_B,
\end{equation}
it is called a product state. Otherwise it is called an entangled state. Thus, we arrive at Definition 1:
\begin{gather*}
    \mbox{\textit{Entangled states are those}}\\
     \mbox{\textit{which are not product states}}. 
\end{gather*}
But this definition fails to offer a profound understanding of entanglement. To simplify the discussion, we focus on finite dimensional Hilbert spaces. Fixing a basis $\{|i\rangle_A\}$ for $H_A$ and a basis $\{|i\rangle_B\}$ for $H_B$, the state $|\psi_{AB}\rangle$ can be written as
\begin{equation}
    |\psi\rangle_{AB} = \sum_{i,j}c_{ij}|i\rangle_A\otimes|j\rangle_B.
\end{equation}
In most cases, the state $|\psi\rangle_{AB}$ is not a product state. For example, the following four Bell states \cite{nielsen2010quantum} are all not product states
\begin{subequations}
\begin{gather}
    |\Phi_+\rangle_{AB} = \frac{1}{\sqrt{2}}\left(|0\rangle_A\otimes|0\rangle_B+|1\rangle_A\otimes|1\rangle_B\right),\\
    |\Psi_+\rangle_{AB} = \frac{1}{\sqrt{2}}\left(|0\rangle_A\otimes|1\rangle_B+|1\rangle_A\otimes|0\rangle_B\right),\\
    |\Phi_-\rangle_{AB}\equiv \frac{1}{\sqrt{2}}\left(|0\rangle_A\otimes|0\rangle_B-|1\rangle_A\otimes|1\rangle_B\right),\\
    |\Psi_-\rangle_{AB}\equiv \frac{1}{\sqrt{2}}\left(|0\rangle_A\otimes|1\rangle_B-|1\rangle_A\otimes|0\rangle_B\right),
\end{gather}
\end{subequations}
and thus all are entangled according to definition 1.

\section{}
For bipartite systems, one can give a more positive definition of entanglement. Recall that a general bipartite quantum state can be written as
\begin{equation}
    |\psi\rangle_{AB}=\sum_{ij}c_{ij}|i\rangle_A\otimes|j\rangle_B.
\end{equation}
The entanglement information is then contained in the matrix $\textbf{M}\equiv (c_{ij})$. One can perform a singular value decomposition $\textbf{M}=\textbf{U}\textbf{D}\textbf{V}$, such that
\begin{equation}
    |\psi\rangle_{AB}=\sum_{i=1}^r\lambda_k|u_k\rangle_A\otimes|v_k\rangle_B,
\end{equation}
where $|u_k\rangle_A\equiv \sum_iu_{ik}|i\rangle_A$, $|v_k\rangle_B\equiv\sum_jv_{kj}|j\rangle_B$, and $\lambda_k\equiv d_{kk}$. This is called the \textbf{Schmidt decomposition} of $|\psi\rangle_{AB}$, where $r$ (the rank of the matrix $\textbf{M}$) is called the \textbf{Schmidt rank} of $|\psi\rangle_{AB}$ \cite{bengtsson2017geometry}. Thus, we arrive at Definition 2:
\begin{gather*}
    \mbox{\textit{Bipartite entangled states are those}}\\ 
    \mbox{\textit{with Schmidt rank greater than one}}. 
\end{gather*}
In this framework, the Bell states $|\Phi_+\rangle_{AB}$ is considered \textbf{maximally entangled} due to its rank being equal to $2$. Unfortunately, Schmidt decomposition does not exist for $N$-partite states where $N$ exceeds 2. However, within the same spirit, it is still possible to express certain multi-partite entangled states. For example, for $N=3$, the following Greenberger–Horne–Zeilinger (GHZ) state is entangled
\begin{align}
    |\mbox{GHZ}\rangle_{ABC} &\equiv \frac{1}{\sqrt{2}}(|0\rangle_A\otimes|0\rangle_B\otimes|0\rangle_C\nonumber\\
    &+|1\rangle_A\otimes|1\rangle_B\otimes|1\rangle_C).
\end{align}

\section{}
For an $N$-partite system, the Hilbert space of $N$ sub-systems with Hilbert spaces $H_1$, $H_2$, $\cdots$, $H_N$, is given by $H_1\otimes H_2\otimes \cdots \otimes H_N$. As a matter of fact, through the application of local unitary transformations on each qubit, entanglement cannot be generated from a product state. For example
\begin{align}
    &(U_1\otimes U_2\otimes\cdots \otimes U_N)|n_1\rangle\otimes |n_2\rangle \otimes \cdots \otimes|n_N\rangle \nonumber\\
    &= |n_1^\prime\rangle \otimes |n_2^\prime\rangle \otimes \cdots \otimes |n_N^\prime\rangle
\end{align}
is still a product state. Thus, we arrive at Definition $3$: 
\begin{gather*}
    \mbox{\textit{Entangled states are equivalence classes }}\\
     \mbox{\textit{under the local unitary transformations,}}\\ 
    \mbox{\textit{with the exclusion of the class of product states}}.
\end{gather*}
Within this framework \cite{grassl1998computing, onishchik2012lie}, entanglement theory is the study of the intrinsic geometric properties of multipartite entangled states. To distinguish different entanglement classes under the action of local unitary group $G=U_1\otimes U_2\otimes \cdots \otimes U_N$, an \textbf{entanglement invariant} becomes necessary \cite{miyake2003classification}. Recall that for the bipartite case, one can use the Schmidt rank to distinguish entangled states from product states. Hence, for a bipartite qubit state $|\psi\rangle_{AB}=\sum_{i,j=0}^1c_{ij}|i\rangle_A\otimes|j\rangle_B$ with a Hilbert space given by $\mathbb{C}^2\otimes\mathbb{C}^2$, the determinant of the matrix $\textbf{M}=(c_{ij})$, defined as $\det\textbf{M} =\det\textbf{D}=\lambda_{1}\lambda_2$ serves as an entanglement invariant. Here, $\textbf{M}$ is invariant under $\mbox{SL}(2,\mathbb{C})^{\otimes 2}$ transformations, and thus invariant under $\mbox{SU}(2,\mathbb{C})^{\otimes 2}$ transformations. Additionally, $\det\textbf{M}$ vanishes when the matrix $\textbf{M}$ is not full rank. In other words, a bipartite qubit state $|\psi\rangle_{AB}$ is a product state only when the discriminant $\det\textbf{M}= c_{00}c_{11}-c_{01}c_{10}$ vanishes. For example, a direct computation yields 
\begin{equation}
    \det(|\Phi_+\rangle_{AB})=1,
\end{equation}
indicating that the Bell state $|\Phi_+\rangle_{AB}$ is entangled. It is noteworthy that the Bell states $|\Phi_-\rangle_{AB}$, $|\Psi_+\rangle_{AB}$ and $|\Psi_-\rangle_{AB}$ can be obtained from $|\Phi_+\rangle_{AB}$ by applying Pauli $Z$ and $X$ gates exclusively on the first qubit
\begin{subequations}
\begin{align}
    &|\Phi_-\rangle_{AB}=(Z\otimes I)|\Phi_+\rangle_{AB},\\
    &|\Psi_+\rangle_{AB}=(X\otimes I)|\Phi_+\rangle_{AB},\\
    &|\Psi_-\rangle_{AB}=(ZX\otimes I)|\Phi_+\rangle_{AB}.
\end{align}
\end{subequations}
This demonstrates that all four Bell states are in the same entanglement class under the action of local unitary transformations. Notice that since $\det(|\Phi_+\rangle_{AB})=\det(|\Psi_+\rangle_{AB})=1$ and $\det(|\Phi_-\rangle_{AB})=\det(|\Psi_-\rangle_{AB})=-1$, the square of the discriminant as a polynomial function of the coefficients $c_{ij}$ serves as a more suitable measure of entanglement.

\section{}
For tri-partite qubit states, there exists a similar entanglement measure called hyper-determinant. For a tri-partite qubit state $|\psi\rangle_{ABC}=\sum_{i,j,k=0}^1c_{ijk}|i\rangle_A\otimes|j\rangle_B\otimes|k\rangle_C$, the hyperdeterminant is defined by
\begin{align}
    &\mbox{Det}(|\psi\rangle_{ABC})\equiv c^2_{000}c^2_{111}+c^2_{001}c^2_{110}+c^2_{010}c^2_{101}\nonumber\\
    &+c^2_{100}c^2_{011}-2c_{000}c_{001}c_{110}c_{111}-2c_{000}c_{010}c_{101}c_{111}\nonumber\\
    &-2c_{000}c_{011}c_{100}c_{111}-2c_{001}c_{010}c_{101}c_{110}\nonumber\\
    &-2c_{001}c_{011}c_{110}c_{100}-2c_{010}c_{011}c_{101}c_{100}\nonumber\\
    &+4c_{000}c_{011}c_{101}c_{110}+4c_{001}c_{010}c_{100}c_{111}.
\end{align}
The hyper-determinant is invariant under $\mbox{SL}(2,\mathbb{C})^{\otimes 3}$ transformations, and thus invariant under $\mbox{SU}(2,\mathbb{C})^{\otimes 3}$ transformations. The utilization of the hyper-determinant demonstrates that the $W$ state, characterized as
\begin{align}
    &|W\rangle_{ABC}\equiv\frac{1}{\sqrt{3}}(|0\rangle_A\otimes|0\rangle_B\otimes|1\rangle_C\nonumber\\
    &+|0\rangle_A\otimes|1\rangle_B\otimes|0\rangle_C+|1\rangle_A\otimes|0\rangle_B\otimes|0\rangle_C)
\end{align}
is in a distinct entanglement class under the action of local unitary group, compared to the GHZ state. A direct computation yields
\begin{equation}
    \mbox{Det}(|\mbox{GHZ}\rangle_{ABC})=\frac{1}{4},\mbox{Det}(|W\rangle_{ABC})=0.
\end{equation}
Thus, the $W$ and GHZ states belong to different entanglement classes according to definition 3.

\section{}
The discussion is not necessarily constrained to qubits. One can also consider multi-partite qutrit states. For a tri-partite qutrit state $|\psi\rangle_{ABC}\equiv \sum_{i,j,k=1,2,3}c_{ijk}|i\rangle_A\otimes|j\rangle_B\otimes|k\rangle_C$, one can derive Nurmiev’s normal form \cite{nurmiev2000orbits} through $\mbox{SU}(3,\mathbb{C})^{\otimes 3}$ transformations
\begin{align}
    &\overline{|\psi\rangle}_{ABC}= a_1(|1\rangle_A\otimes|1\rangle_B\otimes|1\rangle_C+|2\rangle_A\otimes|2\rangle_B\otimes|2\rangle_C\nonumber\\
    &+|3\rangle_A\otimes|3\rangle_B\otimes|3\rangle_C)+a_2(|1\rangle_A\otimes|2\rangle_B\otimes|3\rangle_C\nonumber\\
    &+|2\rangle_A\otimes|3\rangle_B\otimes|1\rangle_C+|3\rangle_A\otimes|1\rangle_B\otimes|2\rangle_C)\nonumber\\
    &+a_3(|1\rangle_A\otimes|3\rangle_B\otimes|2\rangle_C+|2\rangle_A\otimes|1\rangle_B\otimes|3\rangle_C)\nonumber\\
    &+|3\rangle_A\otimes|2\rangle_B\otimes|1\rangle_C).
\end{align}
The entanglement classes are determined by three fundamental invariants $I_6$, $I_9$ and $I_{12}$, which can be explicitly expressed as \cite{olver1999classical}
\begin{subequations}
\begin{align}
I_6&=a_1^6 + a_2^6 + a_3^6 - 10(a_1^3a_2^3 +a_1^3a_3^3 +a_2^3a_3^3),\\
I_9&=-(a_1^3-a_2^3)(a_1^3-a_3^3)(a_2^3-a_3^3),\\
I_{12}&=-(a_1^3+a_2^3+a_3^3)[(a_1^3+a_2^3+a_3^3)^3+(6a_1a_2a_3)^3].
\end{align}
\end{subequations}
The invariants $I_6$ and $I_{12}$ are determined up to nonzero scalar multiples, whereas $I_{12}$ is solely determined up to a scalar multiple of $I_6^2$. Various conventions exist regarding the representation of $I_{12}$. For example, the Bremner invariant, denoted as $J_{12}$, is related to $I_{6}$ and $I_{12}$ by $-I_{12}-I_6^2 = 24J_{12}$. As such, the explicit form
of the $3\times 3\times 3$ hyperdeterminant is given by \cite{bremner20143}
\begin{equation}
\Delta=I_6^3I_9^2-I_6^2J_{12}^2+36I_6I_9^2J_{12}+108I_9^4-32J_{12}^3.
\end{equation}
For instance, the following tri-partite qutrit state is maximally entangled when both $A$ and $B$ are non-zero
\begin{align}
    |\phi&\rangle_{ABC}\equiv \alpha(|3\rangle_A\otimes |2\rangle_B\otimes |1\rangle_C+|1\rangle_A\otimes |2\rangle_B\otimes |3\rangle_C)\nonumber\\
    &+\beta(|3\rangle_A\otimes |1\rangle_B\otimes |2\rangle_C+|1\rangle_A\otimes |3\rangle_B\otimes |2\rangle_C\nonumber\\
    &+|2\rangle_A\otimes |3\rangle_B\otimes |1\rangle_C +|2\rangle_A\otimes |1\rangle_B\otimes |3\rangle_C).
\end{align}
The reason behind this is that the three fundamental invariants $I_6$, $I_9$ and $I_{12}$ for $|\psi_0\rangle$ are $I_6=-8\alpha^2\beta^4$ and $I_9=I_{12}=0$. Thus, the hyperdeterminant $\Delta$ for $|\phi\rangle_{ABC}$ has the form
\begin{equation}
    \Delta(|\phi\rangle_{ABC})=\frac{1}{1728}I_6^6=\frac{4096}{27}(\alpha\beta^2)^{12},
\end{equation}
which is non-zero when both $\alpha$ and $\beta$ are non-zero.

\section{}
For a general multipartite entangled state, if one can identify all invariants under local unitary transformations, the task of classifying entangled states is accomplished \cite{bennett2000exact}. But this turns out to be an extremely difficult task. Fortunately, for multipartite entangled state with certain symmetries, the task is greatly simplified. Notice that all previously discussed important entangled states, like Bell states, GHZ states and $W$ states, are all symmetric with respect to permutation of qubits. In general, $N$-partite permutation symmetric states can be defined as
\begin{equation}
    |\psi\rangle = \frac{1}{\sqrt{N!A_N}}\sum_{\sigma\in S_N}|n_{\sigma(1)}\rangle\otimes \cdots\otimes |n_{\sigma(N)}\rangle,
\end{equation}
\begin{figure}
     \centering
     \begin{subfigure}[htb]{0.22\textwidth}
          \centering
         \includegraphics[width=\linewidth]{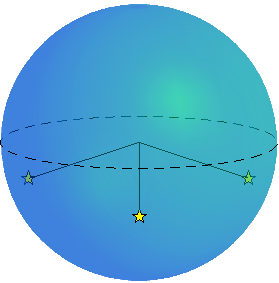}
         \caption{GHZ-like states}
         \label{fig:1}
     \end{subfigure}
     \hfill
     \begin{subfigure}[htb]{0.22\textwidth}
          \centering
         \includegraphics[width=\linewidth]{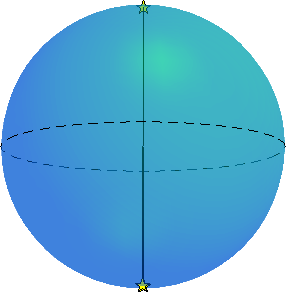}
         \caption{W-like States}
         \label{fig:2}
     \end{subfigure}
    \caption{Majorana representation for tri-partite entangled states \cite{kam2020three}.}
        \label{fig:MajoranaStars}
\end{figure}
where $S_N$ is the symmetric group on $N$ letters, and $A_N$ is a normalization factor. Remarkably, these types of multipartite entangled state which are symmetric respect to permutation of qubits, can be represented geometrically by an assembly of $N$ unordered points on a unit sphere. The reason is simple: a permutation symmetric state, i.e., $\sigma(|\psi\rangle)=|\psi\rangle$ for arbitrary $\sigma\in S_n$, is homeomorphic to an $N$-fold symmetric tensor product of spheres. In physics literature, this is called the \textbf{Majorana representation}, where the unordered points are referred to as \textbf{Majorana stars} \cite{ribeiro2011entanglement}. In Fig.\:\ref{fig:MajoranaStars}, we illustrate the Majorana representation for GHZ-like states and $W$-like states, while the former has three distinct stars and the later has only two distinct stars. Thus, we arrive at Definition 4: 
\begin{gather*}
    \mbox{\textit{For those states which are invariant}}\\
    \mbox{\textit{under permutation of qubits}},\\
    \mbox{\textit{the entangled classes are determined by}}\\ 
    \mbox{\textit{the number of distinct Majorana Stars}}.
\end{gather*} 
When Majorana stars undergo stereographic projection onto the complex plane, the degenerate patterns the stars display align with the nature of roots of a polynomial termed the \textbf{Majorana polynomial}. A non-zero discriminant of the Majorana polynomial corresponds to a maximally entangled state. 

The state with the fewest number of stars is called the \textbf{coherent state} \cite{kam2023coherent}, wherein all $N$ stars degenerate into a single star. It represents the most classical state without any entanglement. In this regard, there exists an onion structure for permutation symmetric states:
\begin{align}
    \mbox{spin-1 coherent states}  &\prec \mbox{Bell states},\nonumber\\
    \mbox{spin-$\frac{3}{2}$ coherent states} &\prec \mbox{$W$ states} \prec \mbox{GHZ states},\nonumber\\
    \vdots\:\:\:\:\:\:\:\:\:\:&\:\:\:\:\:\:\:\:\:\:\:\:\:\:\:\:\:\:\:\:\:\:\:\:\:\:\:\:\:\:\:\:\:\:\vdots\nonumber\\
    \mbox{spin-$\frac{N}{2}$ coherent states}&\prec\cdots \prec \mbox{$N$-qubit GHZ states}.
\end{align}
The above sets up a partial order filtering that always begins with coherent states, which are the most classical, and ends with GHZ states, renowned for being maximally entangled. In this regard, for permutation symmetric states, the problem of classifying entanglement transforms into the spherical design problem, also known as the Thomson problem \cite{bengtsson2017geometry}. 

To conclude, the four definitions of entangled states discussed here are all interconnected. The third definition represents an advancement over the first, as it encompasses a wider range of scenarios, albeit with computational complexities.  Additionally, the fourth definition essentially restates the second one when specifically addresses bipartite symmetric states. 

\bibliography{ExampleRefs.bib}

\end{document}